\documentclass[conference]{IEEEtran}

\usepackage{cite}
\usepackage{graphicx}
\usepackage{amsmath,amssymb}
\usepackage{hyperref}
\usepackage{caption}
\usepackage{subcaption}
\usepackage{booktabs} 
\IEEEoverridecommandlockouts
\begin{document}

\title{Designing Neural Synthesizers for Low-Latency Interaction}
\author{
    \thanks{
    This is the Author’s Accepted Manuscript (AAM), accepted for publication in the \textit{Journal of the Audio Engineering Society}.}
    \IEEEauthorblockN{
        Franco Caspe\IEEEauthorrefmark{1}\thanks{Author to whom correspondence should be addressed, email: f.s.caspe@qmul.ac.uk}, 
        Jordie Shier\IEEEauthorrefmark{1}, 
        Mark Sandler\IEEEauthorrefmark{1}, 
        Charalampos Saitis\IEEEauthorrefmark{1}, 
        Andrew McPherson\IEEEauthorrefmark{2}
    }
    \IEEEauthorblockA{\IEEEauthorrefmark{1}Centre for Digital Music, Queen Mary University of London, London, UK}
    \IEEEauthorblockA{\IEEEauthorrefmark{2}Dyson School of Engineering, Imperial College London, London, UK}
}

\maketitle

\begin{abstract}
Neural Audio Synthesis (NAS) models offer interactive musical control over high-quality, expressive audio generators.
While these models can operate in real-time, they often suffer from high latency, making them unsuitable for intimate musical interaction. The impact of architectural choices in deep learning models on audio latency remains largely unexplored in the NAS literature.
In this work, we investigate the sources of latency and jitter typically found in interactive NAS models. We then apply this analysis to the task of timbre transfer using RAVE, a convolutional variational autoencoder for audio waveforms introduced by Caillon et al. in 2021. Finally, we present an iterative design approach for optimizing latency.
This culminates with a model we call BRAVE (Bravely Realtime Audio Variational autoEncoder), which is low-latency and exhibits better pitch and loudness replication while showing timbre modification capabilities similar to RAVE. We implement it in a specialized inference framework for low-latency, real-time inference and present a proof-of-concept audio plugin compatible with audio signals from musical instruments.
We expect the challenges and guidelines described in this document to support NAS researchers in designing models for low-latency inference from the ground up, enriching the landscape of possibilities for musicians.
\end{abstract}

\section{INTRODUCTION}\label{sec:introduction}

In recent years, interest in interactive Neural Audio Synthesis (NAS) algorithms has increased within the music research community.
NAS algorithms utilize deep neural networks, trained on audio datasets, and recent advancements have paved the way for high-quality synthesis \cite{oord_wavenet_2016}.
The flexibility of these architectures has enabled a variety of control modalities, ranging from MIDI input \cite{narita_ganstrument_2023} to audio-based control \cite{engel_ddsp_2020, caillon_rave_2021}, and has inspired entirely new interfaces \cite{privato_magnetic_2023}.
The use of variational autoencoders \cite{kingma_auto-encoding_2014} for neural audio synthesis has been particularly fruitful, with the RAVE model \cite{caillon_rave_2021} facilitating a range of real-time interactive musical applications and inspiring the development of novel interaction modalities \cite{privato_magnetic_2023, scurto_soundwalking_2023}.

Many NAS algorithms support real-time inference, but they are generally unsuitable for live instrumental performance because they often introduce latencies of several hundred milliseconds \cite{caillon_streamable_2022}.
This is an important factor in music performance; systems that offer low latency can lead to richer and better-perceived experiences \cite{schmid_measuring_2024,morreale_magpick_2019,jack_effect_2016}. 
However, latency in NAS models has yet to be formally investigated, which hinders the development of low-latency musical NAS models. 

In this work, we first present an analysis of primary latency sources in interactive, audio-to-audio NAS algorithms.
These are: 1) buffering delay; 2) cumulative delay; 3) representation delay; 4) data-dependent latency; and 5) positional uncertainty (jitter). As a case study, we analyse RAVE's architecture for these sources of latency, highlighting how this is unsuitable for low-latency interaction. To support this analysis we design an open-source toolkit that we use to empirically evaluate latency during inference.

Building on our initial analysis, we present a design approach where we iteratively modify the RAVE architecture to achieve our target latency and jitter goals for musical interaction, as well as computing requirements for real-time inference on a consumer CPU.

Our iterative design culminates with a model we call BRAVE (Bravely Realtime Audio Variational autoEncoder), which we release in a proof-of-concept audio plugin for musicians to experiment with.
See the accompanying website for training source code, and audio examples  \footnote{\url{https://fcaspe.github.io/brave}}.

We evaluate the model variants on an audio timbre transfer task, i.e., modifying the timbre of an audio input while preserving its musical content, and show that BRAVE improves upon RAVE in preserving musical content (in terms of pitch and loudness) and presents similar capabilities for reproducing the timbre of the training dataset.
The results of our analysis and evaluation point to the importance of the model's receptive field in enabling rich musical affordances such as timbre transfer.
The ability of the decoder to consider time-varying trajectories of latent representations, supported by a suitable receptive field, is critical in the emergence of useful musical interactions.
These findings suggest practical methods for future development of low-latency NAS systems beyond RAVE.

In summary, in this work, we put latency at the center of NAS design.
We provide a design framework that leverages analytical and empirical observations grounded in our musical goals, which motivate and inform architectural decisions, ensuring that both performance and responsiveness are prioritized.
We hope that this perspective, along with the results of this research, will not only support the improvement of existing NAS models but also support the development of novel systems that enrich the landscape of DMIs for musicians.

\section{BACKGROUND}

\subsection{Real-time Constraints for Musical Interaction}

Low action-to-sound latency and timing stability are crucial in DMIs for supporting control intimacy, time-keeping, and developing performance skills and personal style \cite{mcpherson_action-sound_2016}. Wessel and Wright \cite{wessel_problems_2001} suggest an upper bound on latency of 10ms and a jitter (variation of latency) of $\pm$1 ms for musical interaction. Furthermore, Jack et al. \cite{jack_effect_2016} conducted an empirical study that supported this recommendation for percussive instruments, demonstrating that 10ms of latency was acceptable to performers in the absence of jitter; however, 20ms of latency or 10$\pm$3 ms of latency significantly degraded the experience of performance.
Lester and Boley \cite{lester2007effects} performed subjective tests with practicing musicians to investigate the effect of latency under different monitoring scenarios. Their results highlight the context-dependent nature of latency perception with sensitivity amongst musicians varying based on the instrument and type of monitoring device. In general, however, their results reinforce the 10ms bound for minimal impact on musical performance.
In this work, we aim to reach strict latency constraints that approximate those of 10$\pm$1 ms reported in previous work focused on instrumental interaction.

\subsection{ Interactive Neural Audio Synthesis }

NAS is a technique that employs Deep Neural Networks (DNNs) as learned audio synthesizers. 
Neural synthesis has its origins in offline generation; however, increased computation capabilities have enabled real-time inference, which has opened the doors for NAS use in live performance systems with control schemes that span from MIDI inputs \cite{narita_ganstrument_2023,wu_midi-ddsp_2022} to Music Information Retrieval features \cite{pasini_musika_2022}, latent spaces \cite{vigliensoni_steering_2023, esling_generative_2018}, and other audio signals \cite{caillon_rave_2021, pasini_bass_2024}.
Despite this, latency is typically only addressed after models are trained \cite{caillon_streamable_2022}.
A common solution in neural audio plug-in development is to report latency to the audio workstation for it to be compensated by re-aligning other audio tracks, which is incompatible with our live use case.

Recently, a set of NAS models such as RAVE and the Differentiable DSP (DDSP) Decoder~\cite{engel_ddsp_2020} have gained popularity due to their ability to transform audio in real-time according to learned timbral and dynamic characteristics of a dataset. These models can supplement musical instruments with new timbral possibilities, however, such models are not optimized for real-time interaction with the audio of musical instruments, due to their high latency.

Low-latency exceptions to this case are specific models designed for emulating traditional audio effects \cite{carson_sample_2024} and also systems for audio-driven control of traditional synthesizers like the Envelope Learning \cite{caspe_fm_2023}, Timbral Remapping \cite{shier_real-time_2024} and guitar percussive technique \cite{martelloni_real-time_2023} models, designed with low latency interaction with musical instruments in mind. However, these are restricted to specific DSP synthesis techniques such as monophonic FM synthesis, percussive 808-style emulation, or modal synthesis.
In certain contexts, the base latency of NAS models like RAVE may be deemed acceptable,
as in with novel musical interfaces where a fast action-to-sound response is not necessarily expected \cite{privato_sketching_2024}, or in audio plugins for production use \cite{caillon_streamable_2022}.
We focus on audio-driven interactions, specifically timbre transfer, where audio from an instrument is transformed in real-time.
We chose RAVE over DDSP due to its flexibility in handling a broader range of source material, without being constrained to monophonic or harmonic instruments.

\subsection{Real-Time Factor}

Algorithms, including neural nets, are assessed for real-time operation using the Real Time Factor (RTF), defined as the ratio between processing time $t_p$ and the duration of the input $t_i$, $RTF = t_p/t_i$. It follows that a model requires a $RTF < 1$ for a specific block size to process it in real-time with specific hardware.
Low-latency models require processing short audio blocks at a high rate, which limits the available processing time for a single inference pass.

\subsection{Streaming NAS algorithms}

Audio algorithms that support real-time interaction must process sequential blocks of one or more audio samples at a time instead of the entire duration of an input in a single pass. This approach is known as \textit{block processing}. Generative models that support this scheme are called \textit{streaming} models. In this case, block processing has to be supported by all the layers of a model  \cite{caillon_streamable_2022}. Each layer is required to keep track of its internal state between subsequent forward passes to guarantee a continuous output signal between blocks.
For instance, convolutional layers are reconfigured with a \textit{cached padding} mechanism \cite{rybakov_streaming_2020} where the end of the input tensor is retained and used to pad the next input.

The outputs from causal systems do not depend on future inputs. This is a strict requirement for streaming models. Nevertheless, non-causal systems with finite \textit{lookahead}, that is, systems that process a finite number of future input timesteps can be reconfigured for causal operation by introducing delays that allow all required samples to be acquired before processing.
Although non-causal models typically afford better reconstruction quality \cite{caillon_streamable_2022,defossez_high_2022}, in some applications it is desirable to avoid this additional cumulative delay by training in a causal way \cite{steinmetz_efficient_2021,radfar_conmer_2023,martelloni_real-time_2023}.

Designers of streaming models can achieve tolerable latency for several tasks simply by employing causal training and processing relatively short audio blocks. The authors of the EnCodec neural codec \cite{defossez_high_2022} introduced a streaming and causal architecture, reporting a \textit{theoretical minimum latency} of 13 ms, determined by the shortest block supported. However, given the model size (23.3 M parameters), it is unlikely that inference can be performed on a consumer CPU at the frame rate that would support such latency. In a plausible streaming scenario, longer audio windows are accumulated before processing by the coder; this improves its RTF but increases latency, although it remains within tolerable limits for the task.

Similarly, the spoken dialog framework Moshi \cite{defossez_moshi_2024} uses causal models to encode and decode audio frames of 80 ms, performing temporal aggregation in parallel with a causal language model, yielding a final latency of 200 ms. This demonstrates how, for several applications, tolerable latency can be achieved by optimizing specific system components in isolation. Nevertheless, to support instrumental interaction (10 ms total latency, 3 ms jitter), we argue that a more comprehensive analysis of potential latency sources is necessary before design.

\subsection{Timbre Transfer}
Timbre transfer involves transforming a musical audio signal so that timbre is altered while preserving key performance parameters such as melodies, accents, and rhythm.
Most research in timbre transfer is grounded in the classic (but disputed \cite{siedenburg_four_2017}) ANSI definition, which describes timbre as an attribute of auditory sensation that differentiates sounds of the same pitch and loudness \cite{ansi_psychoacustical_1973}. In this context, timbre is typically treated as a global attribute encompassing the sonic quality of a specific instrument.

Recent approaches in timbre transfer, such as Differentiable Digital Signal Processing (DDSP) \cite{engel_ddsp_2020, hayes_review_2024}, build on this by combining explicit pitch and loudness controls with deep learning models that implicitly model timbre through a spectral loss function.
These models have demonstrated real-time timbre transfer capabilities \cite{carney_tone_2021} but remain constrained by their application to monophonic and harmonic instruments.

NAS provides more flexible, source-agnostic methods for modeling musical audio and learning disentangled representations \cite{luo_dismix_2024}.
Engel et al. pioneered an autoencoder architecture \cite{engel_neural_2017}, which enabled interpolations of instrumental timbres.
Mor et al. \cite{mor_universal_2018} and Alinoori et al. \cite{alinoori_music-star_2022} extended this architecture for timbre transfer by using instrument-specific WaveNet decoders and by applying a teacher forcing technique with paired audio data, respectively.
Building on techniques from neural style transfer in the image domain \cite{gatys_neural_2015}, timbre transfer has also been tackled using various architectures, including autoencoders \cite{bitton_modulated_2018, bonnici_timbre_2022, wu_transplayer_2023}, generative adversarial networks (GANs) \cite{huang_timbretron_2019, lu_play_2019}, attention-based methods \cite{jain_att_2020}, vector-quantized variational autoencoders (VQ-VAEs) \cite{bitton_vector-quantized_2020, cifka_self-supervised_2021}, and, more recently, diffusion models \cite{comanducci_timbre_2023, popov_optimal_2023, mancusi_latent_2024}. 

In this work, we focus on timbre transfer using RAVE, an autoencoder-based neural audio synthesis (NAS) architecture. To our knowledge, RAVE is the only NAS-based timbre transfer model to enable real-time, streaming operation; its architecture -- specifically the information bottleneck of its autoencoder design -- and training on datasets with few instruments have enabled timbre transfer affordances. Our exploration centers on how RAVE’s architectural design influences latency and, critically, how this impacts timbre transfer results.

\section{LATENCY IN NEURAL AUDIO SYNTHESIS}

In this section, we first present an analysis of the sources of latency and jitter that can affect interactive NAS models. Then, we set the focus of this work on RAVE, a successful interactive NAS algorithm, and analyse its sources of latency, highlighting how its design decisions hinder its ability to achieve low action-to-sound latency.

\subsection{Latency Sources}

Here, we present the architectural sources of latency which result in a model's delayed response to an input.

\textbf{Buffering}: Real-time audio systems typically work in a pipeline employing a \textit{double buffering} approach, where one audio block is computed while the next is being captured.
In a double buffering scenario, the total output buffering latency is of two block lengths: one for sample acquisition, and one for computing \cite{mcpherson_action-sound_2016}.
Therefore, the block size determines the upper limit for computation time.
This latency source is not visible during model training.

\textbf{Cumulative Delay}
Cumulative delay arises from reconfiguring non-causal convolutional layers for streaming by adding a delay line that the non-causal layers use for looking ahead in time.
This lag, which accumulates through successive layers, is called \textit{cumulative delay}. It can increase latency in non-causally trained models by hundreds of milliseconds \cite{caillon_streamable_2022} compared to the same architecture trained causally.
This delay appears after training, when the model is reconfigured for streaming.
Its final value depends on the configuration of all the layers and can be computed by propagating the delay of the first layer throughout the network. The Pytorch package \verb|cached_conv| presented by Caillon et al. \cite{caillon_streamable_2022} can compute this delay during model building.

\textbf{Representation} We denote representation latency as the number of samples it takes a feature extractor to produce an expected output. NAS models use audio features as input, such as fundamental frequency (F0) and loudness \cite{engel_ddsp_2020}, filter banks such as Pseudo-Quadrature Mirror Filters (PQMF) \cite{caillon_rave_2021}, Short-Time Fourier Transforms \cite{kaneko_istftnet_2022}, or Mel filterbanks \cite{wu1_ddsp-based_2022}, to name a few.
For instance, linear phase Finite Impulse Response (FIR) filterbanks typically exhibit a group delay $D_g$ of approximately half of the filter length $N_f$ \cite{chandra_garai_group_2011}.
On the other hand, other audio feature extractors, such as F0 trackers can exhibit varying latency, depending on their analysis window and the nature of the signal being measured \cite{de_cheveigne_yin_2002}.

\textbf{Data-Dependent Latency} Convolutional layers are FIR filters, and their coefficients depend ultimately on training data. Models based on convolutional networks can exhibit different degrees of latency depending on the learned coefficients. However, this has not been analyzed by previous literature. In Section \ref{sec:design_decisions} we analyze the effect of the training data on the latency, comparing models with the same architecture trained on different datasets.

\subsection{Positional Uncertainty (Jitter)}
Lossy representations computed using block processing encode the position of an event with a temporal resolution of up to a block size. We call this \textit{positional uncertainty}.
In this case, the exact position in samples of an event is lost, and the model's response will depend not only on the input and training method, but also on the relative position of the event within an input window, which generates jitter.
As mentioned earlier, jitter has to be ideally kept under $\pm$ 1 ms to support intimate control with musical instruments. This requirement suggests that smaller block sizes are desirable to minimize jitter.

\subsection{Case study: RAVE}

We address the design of the RAVE model \cite{caillon_rave_2021} because of its extensive application in real-time synthesis and wide interaction possibilities \cite{scurto_soundwalking_2023,privato_magnetic_2023,armitage_explainable_2023}. We base our work on the \verb|v1| version, available at the official repository \footnote{\url{https://github.com/acids-ircam/RAVE}}, and presented in the original paper.
For our analysis, we assume models operating at a sample rate of 44.1 kHz.

RAVE is a convolutional variational autoencoder (VAE) featuring a compressing encoder and decompressing decoder with strided convolutional layers.
For efficiency purposes, it does not work directly with raw audio waveforms and instead encodes and decodes an audio representation downsampled and split into 16 bands using Pseudo Quadrature Mirror Filters (PQMF) \cite{nguyen_near-perfect-reconstruction_1994}.

\textbf{Compression Ratio} The compression ratio determines how many audio samples are compressed into a single latent timestep. A higher compression ratio typically makes audio generation more efficient, with latent vectors representing high-level information \cite{defossez_high_2022,pasini_music2latent_2024}.
The encoder receives a PQMF representation sequence and progressively downsamples it to generate latent timesteps, each one a vector of size 128. We can define the model's compression ratio $C_r =  N_b \cdot \prod_{j=1}^{M} s_{j}$, where $N_b \in \mathbb{N}$ is the number of bands in the PQMF representation and $ S = \{s_n\}_{n=1}^{M} $; $ s_n \in \mathbb{N} $ a sequence of strides along M convolutional blocks in the encoder.
RAVE features $N_b = 16$, $M=4$, and $S = [4, 4, 4, 2]$ and therefore $C_r = 2048$.

The decoder receives a latent sequence and upsamples it through $M$ transposed convolution \cite{zeiler_deconvolutional_2010} upsampling layers with stride configuration $S$, interleaved with $M$ residual stacks.

\textbf{Receptive Field} The encoding and decoding processes have a much larger temporal context than the compression ratio. This context constitutes the \textit{memory} of the system, is denoted \textit{receptive field} \cite{steinmetz_efficient_2021}, which ensures temporal continuity across audio blocks.

We compute the receptive field of RAVE's modules. Using backpropagation, we assess the temporal length of the gradient between inputs and a single output timestep.
The encoder features a receptive field $R_{fe} = 15449$ samples, enabled by the kernel length and striding of its convolutional layers.

On the other hand, RAVE's decoder processes previous latent timesteps due to its dilated residual stacks. Each stack of $K$ residual layers features a dilation configuration $ D = \{d_n\}_{n=1}^{K} $; $ d_n \in \mathbb{N} $ with residual layers becoming progressively more dilated to increase the number of latent timesteps that are taken into account during the upsampling process.
RAVE features $K=3$ residual layers with dilations $D=[1,3,5]$ repeated in $M=4$ residual blocks for a total receptive field $R_{fd}=16$ latent timesteps.
RAVE's total receptive field ($R_f$) can be computed by considering a sliding encoder receptive field $R_{fe}$ that spans across $R_{fd}$ latent space timesteps (Equation \ref{eq:rfield}). This accounts for a total of 46169 audio samples (1.04 s).
\begin{equation}\label{eq:rfield}
R_f = R_{fe} + (R_{fd} - 1) * C_r
\end{equation}

\textbf{Waveform Generation} RAVE generates audio samples with a waveform synthesizer and a noise generator. The waveform synthesizer projects the decoder's output and generates a loudness envelope and multiband signal (with \textit{tanh} activation) that is inverted back into 2048 samples using PQMF.
The noise generator is based on a strided convolutional stack that re-compresses the decoder's output using a ratio of 1024 and then performs Fast Fourier Transform (FFT) convolution on white noise to generate another multiband signal that is summed with the waveform synthesizer before PQMF inversion.

\textbf{Training }Models are first trained using \textit{representation learning}, which uses a multi-resolution spectral reconstruction loss as a metric to learn a suitable compressed representation, and later with an \textit{adversarial fine-tuning} to improve the decoder's audio quality.
The noise generator is only active during adversarial fine-tuning tuning. This "slightly increases the reconstruction naturalness of noisy signals" \cite[p. 6]{caillon_rave_2021}.
Finally, RAVE can be trained in a \textit{causal} or \textit{non-causal} fashion. Non-causal training allows the model to look ahead half of its receptive field into the future. This increases reconstruction quality at the expense of additional delay. On the other hand, a causal training procedure ensures that the model produces output only by accounting for current and past input. The original RAVE model has 17.6 million trainable parameters.

\subsection{Sources of latency in RAVE}

In this section, we offer a critical analysis of the latency sources present in RAVE's architecture.

\textbf{Buffering Delay}: determined by two audio blocks of 2048 samples, equalling a total of 4096 samples (92.9 ms). 

\textbf{Representation Delay:} The PQMF module features a minimum inter-band attenuation of 100 dB, implemented with 16 polyphase FIR filters for encoding and 16 for decoding. This requires filter lengths of 513 and 33 for the encoding and decoding process respectively. This yields a group delay of 256 samples for encoding and 16 samples for decoding. However, each sample at the decoder is then up-sampled by the number of channels. This yields an equivalent of 256 samples of delay at the decoder process, for a total of 512 samples of representation delay (12 ms).

\textbf{Cumulative Delay:} It manifests when enabling models for streaming that were trained using a \textit{non-causal} approach such as the original RAVE model. We employ the \verb|cached_conv| package to measure the theoretical cumulative delay in both encoder and decoder and find that it corresponds to a total of 566 ms at 44.1kHz, which is enough to cover RAVE's half-second lookahead.

\textbf{Jitter:} RAVE's encoder acts as a learned feature extractor that processes input audio in a block-by-block fashion. Using sliding blocks would break the temporal coherency of the latent space, which, during training, encodes an input signal without overlap.
Because of this, streaming RAVE has to operate with a minimum block size of the compression ratio $C_r = 2048$ samples, which is compressed into a single latent vector timestep of size 128. This can heavily restrict an event's temporal resolution, resulting in jitter when decoding due to the positional uncertainty of the latent representation. For example, consider a trained model that requires an incoming event to be fully captured by the input to decode it. In this case, the jitter could easily cover the range of an audio block, $\pm$ 23.2ms, for a total jitter span of 46.4ms.

\textbf{Summary:} RAVE's design decisions prioritize reconstruction quality at an elevated compression ratio, which is useful for latent space manipulation and automatic synthesis through \textit{prior} networks \cite{caillon_rave_2021}. 

Considering our goal of interaction with musical instruments, it is unclear what architectural features enable RAVE's timbre transfer capabilities. A valid hypothesis could be that a high compression ratio affords a high-level representation of input, capturing complex temporal structure, and allowing the system to learn to track melodies and harmonies. However, RAVE's long receptive field could also aggregate such information.

Interestingly, the features that make this system attractive for automatic music generation make them highly unsuitable for low-latency interaction: a high compression ratio, present in both causal and non-causal versions, introduces unacceptable buffering delay and jitter but enables efficient inference with prior networks, whereas a non-causal mode of operation introduces cumulative delay but improves quality and does not affect offline generation. In the next section, we reverse the design goals, aiming to develop a system for low-latency musical performance.


\section{ARCHITECTURE RE-DESIGN}\label{sec:design_decisions}

RAVE's auto-encoding architecture works well for the timbre transfer case, but cannot support low-latency inference.
This section discusses RAVE's design choices and examines their effect on real-time factor (RTF) and \textit{sound-to-sound latency}. The latter is closely related to action-to-sound latency, representing the delay between a captured sound-producing action performed on an instrument and the model's resulting output.
We perform this analysis by systematically and progressively modifying RAVE's architecture towards a low-latency variant.

\begin{figure*}
\centering
\includegraphics[width=0.9\textwidth]{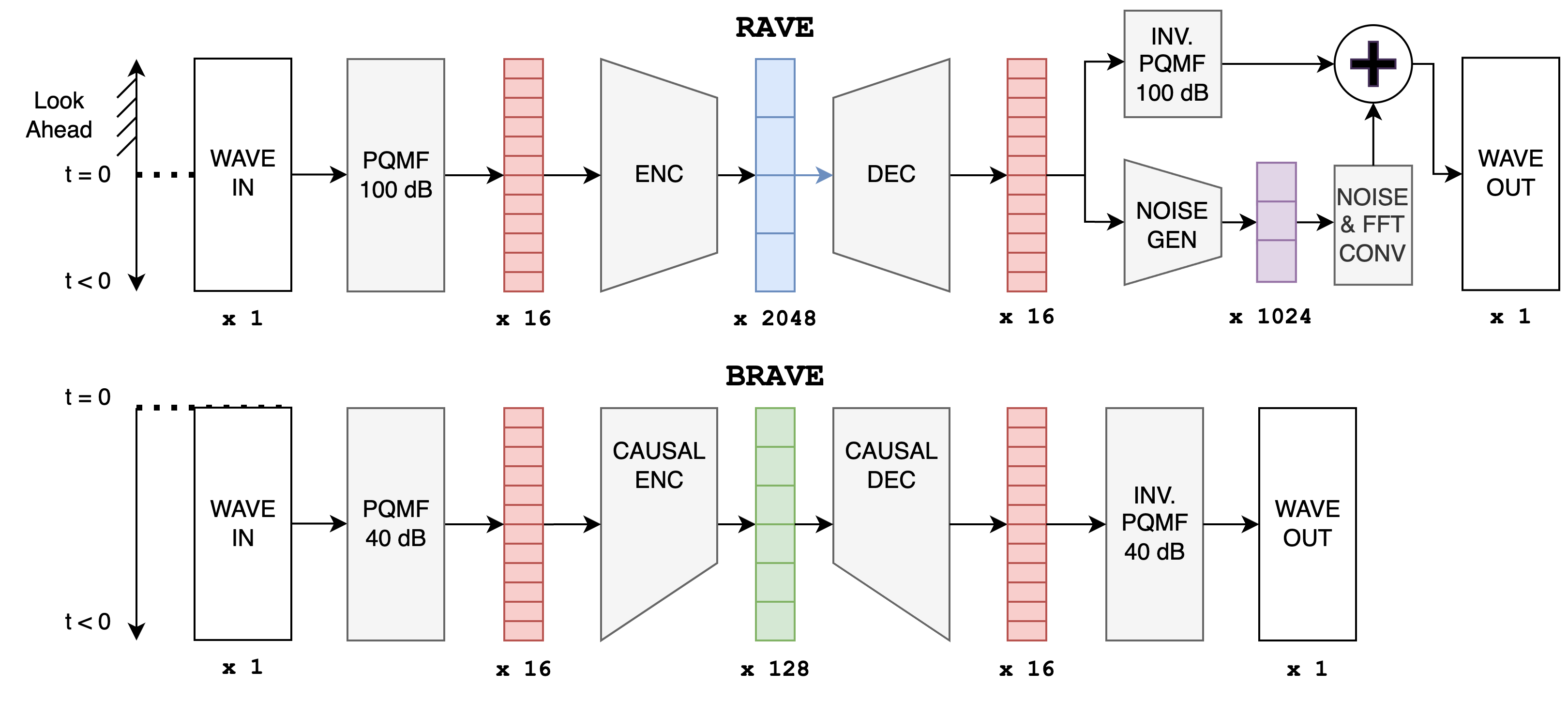}
\caption{Simplified architectural comparison. BRAVE achieves adequate latency ($<10$ ms) and jitter (~3 ms) by removing RAVE's noise generator and using a smaller encoder compression ratio, PQMF attenuation, and causal training, reducing its buffering, representation, and cumulative delays respectively. The number of parameters is also reduced to improve its RTF (see Table \ref{tab:models} ). Numbers in \textbf{\texttt{monospace}} denote the compression ratio of intermediate results.}\label{fig:architecture}
\end{figure*}

Firstly, we present the datasets we use for training the design iterations. Then, we introduce an empirical latency and RTF measurement strategy we use to guide our design space exploration. 
Next, we train variations of RAVE with different $C_r$, \textit{PQMF attenuation}, and $R_f$. This allows us to assess their impact on latency and RTF.
We close this section by presenting BRAVE, a model suitable for low-latency sound-to-sound interaction, which achieves adequate latency $< 10$ms and low total jitter of about 3 ms. Figure \ref{fig:architecture} shows its architecture compared to RAVE's.

\subsection{Training details and Datasets}

We train all models in a \textit{causal} approach to minimize the cumulative delay, following RAVE's two-phase training schedule over a total of 1.5M steps; 1M steps of representation learning and 500k steps of adversarial training with the original RAVE's adversarial model. We train models that encode and decode a single audio channel. Furthermore, we remove RAVE's noise generator from all models except our RAVE reference, because of its independent compression ratio which should be optimized separately, obscuring the analysis of latency and its sources.

We train our different variations of RAVE using percussive and harmonic datasets to analyze whether training data affects the models' latency.
\textbf{Filosax}: An excerpt from the Filosax dataset \cite{foster_filosax_2021} containing 4.5h of solo saxophone, extracted from "Participant 1". We hold the last four takes of the same participant and use them as the test set.
\textbf{Drumset}: A subset of the Expanded Groove MIDI Dataset \cite{callender_improving_2020}. Two hours and 50 minutes of a drummer performing on an electronic drum kit. The selected subset comprises ``drummer1" playing the ``Acoustic kit" and we use samples from the denoted training set. This dataset includes separated test samples which we employ for the test set.

We compute the average loudness of each dataset according to ITU-R BS.1770-4 across all instances, obtaining -18.5 and -36.0 LUFS for \textbf{Filosax} and \textbf{Drumset} respectively. We use this measurement to balance loudness on the test datasets described in Section \ref{sec:testsets}. All datasets are sampled at 44.1 kHz. We do not apply any postprocessing to avoid altering the instruments' relationship between timbre and loudness.

\subsection{Latency and RTF Evaluation}
To guide our design exploration, we empirically measure latency and jitter by testing trained models with a large set of input signals. We measure the system response time as determined by the difference between input and output onset.
We also measure RTF by collecting the time taken for multiple forward passes using different block sizes.
We release this evaluation methodology as an open-source Python package to support NAS developers (see accompanying website).

\textbf{Synthetic Excitation Signals:} 
By synthesizing different excitation signals we can precisely control the onset times, and generate a wide range of spectral characteristics, to assess the models' response under diverse circumstances.
We include white noise, a \textit{dense sinusoidal} containing frequencies increasing logarithmically by a half semitone filling a randomly selected bandwidth, and a harmonic signal with a random pitch and number of harmonics.
All signals are enveloped with an exponential function that decays the amplitude by 60dB over the duration of the signal.

Once signals are generated, they are front-padded with a random duration of silence $N_s\sim\mathcal{U}\{2048, ..., 44100\}$ and are back-padded with a second of silence. A full grid search over signal type, length $N_l \in \{4096, 44100\}$, and amplitude $A \in \{0\text{dB}, -6\text{dB}\}$ (equalling 12 different signals) was conducted for each model and each test repeated 500 times.
All presented results were selected using the signal that leads to the best average latency results for each model and dataset, representing a best-case scenario.

\textbf{Measuring Latency: }
We set the models for streaming operation which introduces cumulative delay. 
Then, latency is measured as the delay between the start of the excitation signal and the first onset measured in the output signal.
Our goal is to measure the time required for a model to respond to an input.
We use two different onset detectors and report the first detected onset between the two. 
The first is a time-domain method, based on the \textit{AmpGate} algorithm implemented in the FluCoMa library \cite{tremblay_fluid_2022}, which responds quickly to percussive onsets, and the second is the spectral-flux method proposed by \cite{bock_maximum_2013}, which is more robust to different signal characteristics such as slower attacks observed in wind instruments.

We obtain the latency and jitter in samples and convert them to milliseconds using the datasets' sampling rate (44.1 kHz). To this value, we add the required buffering delay, equal to two blocks of size equal to the respective compression ratio, therefore showing the minimum sound-to-sound action the model can exhibit. 
While the error range of the onset detectors may vary depending on the data the models render, impacting the jitter results, we use our latency test as a guide for our designs; even though we cannot guarantee the informed jitter ranges to be exact, we expect models with better jitter to show so in the results.

\textbf{Measuring RTF: }
The RTF measurement process starts by processing the Pytorch models into scripted compute graphs (i.e., the sequence of blocks and layers of a model) using the tools provided in the RAVE repository. The scripting process generates a TorchScript \footnote{\url{https://pytorch.org/docs/stable/jit.html}} file that can be loaded in Python or C++ using Libtorch, Torch's C++ API. Once loaded, the models are compiled just in time before execution, optimizing the inference process and improving their RTF in comparison with a standard Pytorch inference pass. Scripted models have been integrated into neural audio plugins without additional optimizations \cite{diaz_interactive_2023,caspe_fm_2023-1,caillon_streamable_2022}. Therefore,
we consider this a faithful approximation of performance in a typical real-time use case.

For this process, we enable cached convolutions in all models to ensure they support streaming operations. However, we skip the latent projection layers the original tool appends to the models' compute graph because of their additional computational requirements. This does not modify the models' performance and serves for latent space manipulation which is outside of the scope of this work.

We design a basic C++ application that loads the scripted models and times their execution time for different audio block sizes. The models are executed 1100 times, with the first 100 executions employed as a "warm-up" for the operating system scheduler to find an optimal way to handle the inference workload, and the rest registered for computing statistics. Then, our application calculates the RTF of each run taking 44.1 kHz as a reference sampling rate and reports the mean and standard deviation. We perform all measurements on a MacBook with an M1 Pro CPU and 16 GB RAM, running MacOS Sonoma 14.3.

\subsection{Compression Ratio Latency}

In this subsection, we empirically analyze the effects of compression ratio in the model, which directly impacts buffering latency and jitter. We train RAVE versions that operate at different compression ratios, without altering other model features such as PQMF attenuation, receptive field, and number of parameters.

We reduce the compression ratio of models by modifying the strides at both the encoder and decoder. However, this also affects directly their receptive field. Since the generator has a fixed receptive field of latent vectors, halving the number of samples encoded in each latent vector halves the encoder's receptive field. We compensate for this by increasing the generator's receptive field. 

We modify the generator's residual blocks' dilations, manipulating $D$. Since the residual layers feature a kernel size of 5, we chose to exponentially increase the dilation factor by a maximum rate of 3, this avoids holes in the decoder's receptive field.
Furthermore, we also modify $K$ in models with small compression ratios, adding more layers to the generator's residual block. This allows us to work with residual blocks with higher dilations at a relatively small increase in the total number of parameters. In doing so, we test models with progressively smaller compression ratios but with similar sizes and receptive fields. 

Following this approach we design models by progressively halving the original compression ratio down to 128. This last model can operate with a small buffering delay of 5.8 ms at a jitter of 2.9 ms, or $\pm$ 1.45 ms which we deem close enough to our initial latency and jitter targets. However, other sources of latency are still present. Please refer to Table \ref{tab:models} for their architecture characteristics.

We train RAVE v1 and all model variations on both datasets.  We observe that all training processes converge similarly.
Latency tests are run against all the trained models and results are shown in Table \ref{tab:latency_1}.  
We observe that latency improves greatly with causal training, and drastically improves with smaller compression ratios. As expected, the model iteration \verb|c128_r10|, with its compression ratio of 128 can provide an acceptable jitter; however, results vary with the training dataset, and all models exceed the 10 ms latency requirement.

\begin{table}[ht]\centering
\scriptsize{
\begin{tabular}{lll}
\hline
& Filosax & Drumset \\
\hline
RAVE v1 & 244.83 [136.60] & 439.89 [84.56] \\
\verb|c2048_r10| & 144.44 [116.28] & 130.62 [42.97] \\
\verb|c1024_r10| & 82.56 [47.05] & 70.40 [41.36] \\
\verb|c512_r10| & 43.42 [21.16] & 39.80 [12.65] \\
\verb|c256_r10| & 30.63 [28.80] & 25.77 [5.71] \\
\verb|c128_r10| & 20.50 [8.34] & 18.50 [3.27] \\
\hline
\end{tabular}
} 
\caption{Latency evaluation for different compression ratios. Reported as the \textit{best average latency} [best jitter] in milliseconds.}\label{tab:latency_1}
\end{table}

\subsection{Multiband Decomposition Latency}

Having lowered the latency and achieved an acceptable jitter, we turn to the PQMF multi-band decomposition module to find other sources of latency we can tune. It is possible to reduce the group delay $D_g$ of the filters by relaxing the inter-band attenuation requirements, obtaining shorter FIR filters with shorter delay.

Although greater inter-band attenuation is typically desired in DSP applications, leading to a better reconstruction with less sub-band aliasing, the impact is less clear in NAS where the leakage between sub-bands could be leveraged by the model or compensated for during training.

We train variations of \verb|c128_r10| with attenuations of 70 and 40 dB, which we denote \verb|c128_r10_p70| and \verb|c128_r10_p40|. They feature PQMF group delays of 256 and 128 samples respectively, while featuring the same $C_r$ and $R_f$ of \verb|c128_r10|. Their architectural description is shown in Table \ref{tab:models}. 
Next, we evaluate their latency response. Results are shown in Table \ref{tab:latency_2}, where we observe a reduction in latency with less PQMF attenuation. We have nearly reached our latency target with the \verb|c128_r10_p40| model, which exhibits latencies around 10 ms, though it shows slightly higher-than-desired jitter values. Nonetheless, we consider these results suitably close to our goal to consider it for real-time implementation.

\subsection{Optimizing for RTF}

Real-time execution can be exceptionally challenging for low-latency models where inference is performed many times per second, i.e., at a high inference frame rate. This is partially due to the increasing number of scaffolding operations conducted by the inference framework such as memory allocation for inputs, outputs, and intermediate results \cite{chowdhury_rtneural_2021}: a model can exhibit a higher RTF if the inference is computed at a smaller block size. On the other hand, the lower $C_r$ models have a similar number of weights compared to the original RAVE, but generate smaller audio blocks at each pass, meaning it is less efficient in terms of computing requirements per sample.

Our initial low-latency model, \verb|c128_r10_p40| cannot run in real-time on our reference CPU. We set out then to reduce its computational requirements. Firstly, we remove the layers with higher dilation in the generator, effectively halving its receptive field. We call this model \verb|c128_r05_p40|.
Next, we halve the hidden sizes of both the encoder and decoder to achieve a model roughly one-third the original size, which can be computed at a suitable RTF. In line with this, we also halve the number of channels of its discriminator. We denote this model BRAVE. See Table \ref{tab:models} for architectural details.

\begin{table*}[h]
\scriptsize{
\begin{tabular}{llllllll}
\hline
\textbf{Model}  & \textbf{Hidden Sizes}   & \textbf{$S$: Strides} & \textbf{$D$: Dilations} & \textbf{PQMF Att. (dB)} & \textbf{$C_r$: C. Ratios} & \textbf{$R_f$: Rec. Field (ms)} & \textbf{\# Parameters (M)}\\
\hline
RAVE v1  \scriptsize{(non causal)}        & {[}64, 128, 256, 512{]} & {[}4, 4, 4, 2{]} & {[}1,3,5{]}                       & 100                     & 2048              & 1047  & 17.6                             \\
\verb|c2048_r10|      & {[}64, 128, 256, 512{]} & {[}4, 4, 4, 2{]} & {[}1,3,5{]}                       & 100                     & 2048              & 1047  & 17.5                             \\
\verb|c1024_r10|      & {[}64, 128, 256, 512{]} & {[}4, 4, 2, 2{]} & {[}3,9,27{]}                      & 100                     & 1024              & 1070   & 16.9                            \\
\verb|c512_r10|       & {[}64, 128, 256, 512{]} & {[}4, 2, 2, 2{]} & {[}3,9,18,36{]}                   & 100                     & 512               & 960  & 18.4                             \\
\verb|c256_r10|       & {[}64, 128, 256, 512{]} & {[}2, 2, 2, 2{]} & {[}3,9,27,36{]}                   & 100                     & 256               & 973  & 16.2                            \\
\verb|c128_r10|       & {[}64, 128, 256, 512{]} & {[}2,2,2,1{]}    & {[}3,9,27,45,63{]}                & 100                     & 128               & 955   & 17.3                           \\
\verb|c128_r10_p70|  & {[}64, 128, 256, 512{]} & {[}2,2,2,1{]}    & {[}3,9,27,45,63{]}                & 70                      & 128               & 947   & 17.3                            \\
\verb|c128_r10_p40|  & {[}64, 128, 256, 512{]} & {[}2,2,2,1{]}    & {[}3,9,27,45,63{]}                & 40                      & 128               & 941  & 17.3                             \\
\verb|c128_r05_p40|  & {[}64, 128, 256, 512{]} & {[}2,2,2,1{]}    & {[}3,9,27,36{]}                   & 40                      & 128               & 517  & 15.2                             \\
BRAVE           & {[}32. 64, 128, 256{]}  & {[}2,2,2,1{]}    & {[}3,9,27,36{]}                   & 40                      & 128               & 517   & 4.9                           \\
\hline
\end{tabular}}
\caption{Summary of models implemented. All models have a latent vector size of 128. All of them are causal and do not have a noise generator, with exception of RAVE. The receptive field assumes a sample rate of 44.1 kHz.}\label{tab:models}
\end{table*}

We evaluate the lightweight models' latency, shown in Table \ref{tab:latency_2} finding that the modifications do not substantially impact latency. Then, we run our RTF test on these models. Results are shown in Table \ref{tab:rtf_1}. We observe that all models drastically improve their RTF with larger block sizes. For instance, BRAVE can be run in real-time but only at a larger block size of 256, which increases buffering latency. We suspect this is due to scaffolding operations within scripted models that require a fixed time, such as memory allocation \cite{ackva_anira_2024}. This prompts us to evaluate an implementation using different tools to avoid that temporal cost.

\begin{table}[ht]\centering
\scriptsize{
\begin{tabular}{llll}
\hline
& Attenuation (dB) & Filosax & Drumset \\
\hline
\verb|c128_r10|     &  100 & 20.50 [8.34] & 18.50 [3.27] \\
\verb|c128_r10_p70| &   70 & 13.58 [10.86] & 13.19 [4.22] \\
\verb|c128_r10_p40| &   40 & 10.46 [6.30] & 9.92 [3.36] \\
\verb|c128_r05_p40| &   40 & 10.22 [7.57] & 9.67 [4.31] \\
BRAVE & 40 & 10.08 [7.80] & 9.75 [2.47] \\
\hline
\end{tabular}
} 
\caption{Latency evaluation for different PQMF attenuations. All modes have the same $C_r$. Reported as the \textit{best average latency} [best jitter] in milliseconds. \texttt{c128\symbol{95}r05\symbol{95}p40} and BRAVE are optimized for RTF.}\label{tab:latency_2}
\end{table}

\begin{table}[ht]\centering
\scriptsize{
\begin{tabular}{lllll}
\hline
Model & 128 & 256 & 512 & 2048 \\
\hline
RAVE v1 & - & - & - & 0.16 (0.02) \\
\verb|c2048_r10| & - & - & - & 0.15 (0.02) \\
\verb|c128_r10| & 2.62 (0.07) & 1.43 (0.03) & 0.74 (0.05) & 0.24 (0.04) \\
\verb|c128_r10_p70| & 2.61 (0.06) & 1.42 (0.03) & 0.73 (0.02) & 0.24 (0.05) \\
\verb|c128_r10_p40| & 2.60 (0.06) & 1.42 (0.03) & 0.73 (0.02) & 0.24 (0.04) \\
\verb|c128_r05_p40| & 2.20 (0.06) & 1.22 (0.03) & 0.62 (0.01) & 0.20 (0.04) \\
BRAVE & 1.18 (0.07) & 0.65 (0.03) & 0.37 (0.08) & 0.10 (0.003) \\
\hline
\end{tabular}
} 
\caption{RTF for selected block sizes. Shown as mean (std).}\label{tab:rtf_1}
\end{table}

\subsection{Low Latency Implementation}\label{sec:bravecpp}

We present a custom C++ inference developed by the first author to support the BRAVE autoencoding architecture and some variations. 
The engine is based on RTNeural \cite{chowdhury_rtneural_2021}, a DNN inference library for real-time DNN inference for audio applications such as audio plug-ins.
We choose RTNeural for its design, which transparently supports causal, block-based inference, prompting the designer to think of the model inference process as a streaming operation. Furthermore, it pre-allocates all memory before inference; this is critical for working at the high frame rates required for low latency.
We extend RTNeural to support strided and transposed convolutions and implement all model components.

We implement three different causal models in our architecture: a version of RAVE v1 without noise generator (\verb|c2048_r10|), a low latency and high-capacity model (\verb|c128_r05_p40|), and BRAVE. 
We achieve the fastest implementation in our reference M1 Pro CPU with the Standard Template Library backend of RTNeural, using Apple Clang 15. Table \ref{tab:rtf_3} compares the RTF between the RTNeural and Libtorch implementations, showing that the latter can execute BRAVE comfortably at the lowest block size and latency, unlike Libtorch. However, it does not improve its RTF with longer block sizes for any of the models, with Libtorch becoming a more efficient option for longer block sizes. We suspect Libtorch may employ computing algorithms optimized for different block sizes. We deploy our models in a proof-of-concept audio plugin (see the accompanying webpage).

\begin{table*}[ht]\centering
\scriptsize{
\begin{tabular}{lllll}
\hline
Model & 128 & 256 & 512 & 2048 \\
\hline
\verb|c2048_r10| & - & - & - & 0.15 (0.02) \\
\verb|c2048_r10| $\dagger$ & - & - & - & 0.30 (0.007) \\
\verb|c128_r05_p40| & 2.20 (0.06) & 1.22 (0.03) & 0.62 (0.01) & 0.20 (0.04) \\
\verb|c128_r05_p40| $\dagger$  & 1.02 (0.005) & 1.01 (0.010) & 1.01 (0.029) & 1.00 (0.008) \\
BRAVE & 1.18 (0.07) & 0.65 (0.03) & 0.37 (0.08) & 0.10 (0.003) \\
BRAVE $\dagger$ & 0.29 (0.006) & 0.29 (0.005) & 0.29 (0.064) & 0.29 (0.020) \\
\hline
\end{tabular}
} 
\caption{RTF comparison of Libtorch and RTNeural (denoted with $\dagger$) models at different block sizes.}\label{tab:rtf_3}
\end{table*}

\section{EVALUATION}
In this section, we set out to determine whether our design variations can support timbre transfer, testing (1) the models' audio quality, (2) their timbre modification, and (3) their content preservation capabilities,  comparing them to RAVE's. We select models that progressively accumulate modifications to assess how causality (\verb|c2048_r10|), compression ratio (\verb|c128_r10|), PQMF attenuation (\verb|c128_r10_p40|), receptive field (\verb|c128_r05_p40|), and model capacity (BRAVE) affect timbre transfer.

Firstly, we assess their audio quality using a simple resynthesis task. Then, we present the datasets we employ as sources for timbre transfer. Next, we evaluate whether a high compression ratio is necessary to perform timbre transfer or it can be achieved using a small compression ratio but a sizable receptive field. To this end, we introduce our Maximum Mean Discrepancy (MMD) test, similar to that of Bitton et al. \cite{bitton_modulated_2018}, designed to quantify differences in timbre distributions between the original and transferred audios.
Finally, we investigate if working with smaller block sizes can improve the temporal resolution of rendered musical events. We compare the models' capabilities to follow musical dynamics and melodies in timbre transfer by evaluating loudness curves and fundamental frequency rendering.

\subsection{Audio Quality Assessment}\label{sec:audioquality}

We employ the widely used Fr\'echet Audio Distance (FAD) \cite{kilgour_frechet_2019} to evaluate audio quality using the VGGish model. 
We compute the background embeddings using both the \textbf{Drumset} and \textbf{Filosax} training datasets. Evaluation embeddings are computed on audio from their corresponding test datasets, which have been resynthesized by each model variant.
We add the original test set as an additional evaluation embedding for reference.
FAD is computed between background and evaluation embeddings for each model variant.
Results are shown in Table  \ref{tab:fad}.

We observe all the models score similarly on their respective datasets, which indicates similar audio quality, with a slight degradation in BRAVE possibly due to its smaller capacity. The notable exception is the high compression ratio models trained on \textbf{Filosax}. 
We suspect such variations in score are due to unstable adversarial training, which negatively impacts the capability of the models to replicate pitched sounds.
This suggests that such models may require larger datasets to stabilize training, as evidenced by the size of datasets used in the original RAVE implementation which are about one order of magnitude bigger than ours.
Please refer to the accompanying website for audio examples.

\begin{table}[ht]\centering
\scriptsize{
\begin{tabular}{lll}
\hline
 & Filosax & Drumset \\
\hline
RAVE v1 & 43.95 & 1.48 \\
\verb|c2048_r10| & 41.54 & 1.55 \\
\verb|c128_r10| & 6.99 & 1.19 \\
\verb|c128_r10_p40| & 7.58 & 1.11 \\
\verb|c128_r05_p40| & \textbf{6.04} & \textbf{1.10} \\
BRAVE & 9.03 & 2.21 \\
\textit{Test Set} & \textit{0.22} & \textit{0.28} \\
\hline
\end{tabular}
\caption{FAD computed on resynthesis of test set.}\label{tab:fad}
} 
\end{table}

\subsection{Testing Datasets}\label{sec:testsets}
We select audio corpora for testing the timbre transfer capabilities of the models. We employ different source sets depending on whether our models were trained on \textbf{Filosax} (melodic test set) or \textbf{Drumset} (percussive test set).

\textit{Melodic test set}: \textbf{Viola}: 26 min of solo viola recordings extracted from the URMP dataset \cite{li_urmp_2019}, and \textbf{Svoice}: 16.5 min of female singing: \textit{participant 01} and \textit{participant 12} \cite{black_automatic_2014}. Each track of these datasets is loudness-normalized to -18 LUFS, to match the \textbf{Filosax} dataset loudness.

\textit{Percussive test set}: \textbf{Candombe}: presented in Nunes et al. \cite{nunes_beat_2015}, we select a single recording for each artist totaling 19 min. \textbf{Beatbox}: we employ the complete Amateur Vocal Percussion Dataset \cite{delgado_amateur_2019}, totalling 18 min. Each track is loudness normalized to -36 LUFS, to match that of the \textbf{Drumset} dataset.

\subsection{Timbre Transfer Evaluation}
We seek a numerical analysis of timbre transfer performance to compare the capabilities of our model to the original RAVE model.
As ground truth audio signals are not available in this evaluation, we consider distributions of timbre features and view timbre transfer as transforming from one distribution to another \cite{moliner_gaussian_2024}.
We turn to the \textit{maximum mean discrepancy} (MMD) \cite{gretton_kernel_2012}, a two-sample statistical test to determine whether samples are drawn from different distributions, which has been successfully applied to evaluate audio timbre transfer \cite{bitton_modulated_2018}.

Given independent samples \( \{ x_i \}_{i=1}^n \) from random variable $X$ following distribution $P$, $ X \sim P $, and \( \{ y_j \}_{j=1}^m \) from $ Y \sim Q $, the unbiased empirical estimator of the squared Maximum Mean Discrepancy (MMD) is:

\begin{align*}
\text{MMD}_u^2[X, Y] = &\ \frac{1}{n(n-1)} \sum_{i=1}^{n} \sum_{\substack{j=1 \\ j \neq i}}^{n} k(x_i, x_j) \\
& + \frac{1}{m(m-1)} \sum_{i=1}^{m} \sum_{\substack{j=1 \\ j \neq i}}^{m} k(y_i, y_j) \\
& - \frac{2}{nm} \sum_{i=1}^{n} \sum_{j=1}^{m} k(x_i, y_j)
\end{align*}

\noindent where $k$ is a Gaussian kernel $k(x,x^{\prime}) = \text{exp}(-\frac{1}{2\sigma}\lVert{x-x^\prime}\rVert^{2})$ and $\sigma$ is the bandwidth parameter.
We select $\sigma$ independently for each test as the median $L_2$ distance between all pairwise sample combinations \cite{gretton_kernel_2012}.

Timbre distributions are generated for each audio corpus using Mel-frequency cepstral coefficients (MFCCs). 
MFCCs are computed with FFTs of size 2048 with 75\% overlap and 128 Mel-bands; MFCCs 2-13 are then selected \cite{cifka_self-supervised_2021}.
Texture windows are created from MFCC frames by average pooling using a window of 40 frames ($\approx$0.5s) with 50\% overlap.
The set of texture windows for each audio corpora is used for MMD.

\begin{figure*}[ht]
    \centering
    \begin{subfigure}[b]{0.49\textwidth}
        \centering
        \includegraphics[width=\textwidth]{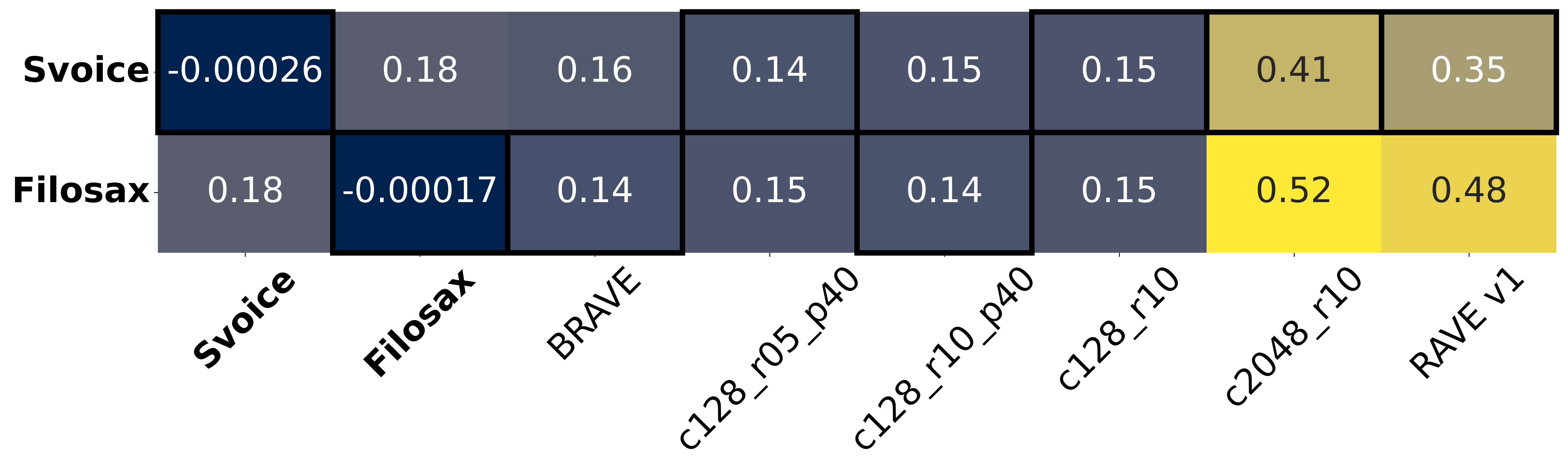}
        \caption*{\textbf{Svoice} processed by models trained on \textbf{Filosax}}
        \label{fig:sub1}
    \end{subfigure}
    \hfill
    \begin{subfigure}[b]{0.49\textwidth}
        \centering
        \includegraphics[width=\textwidth]{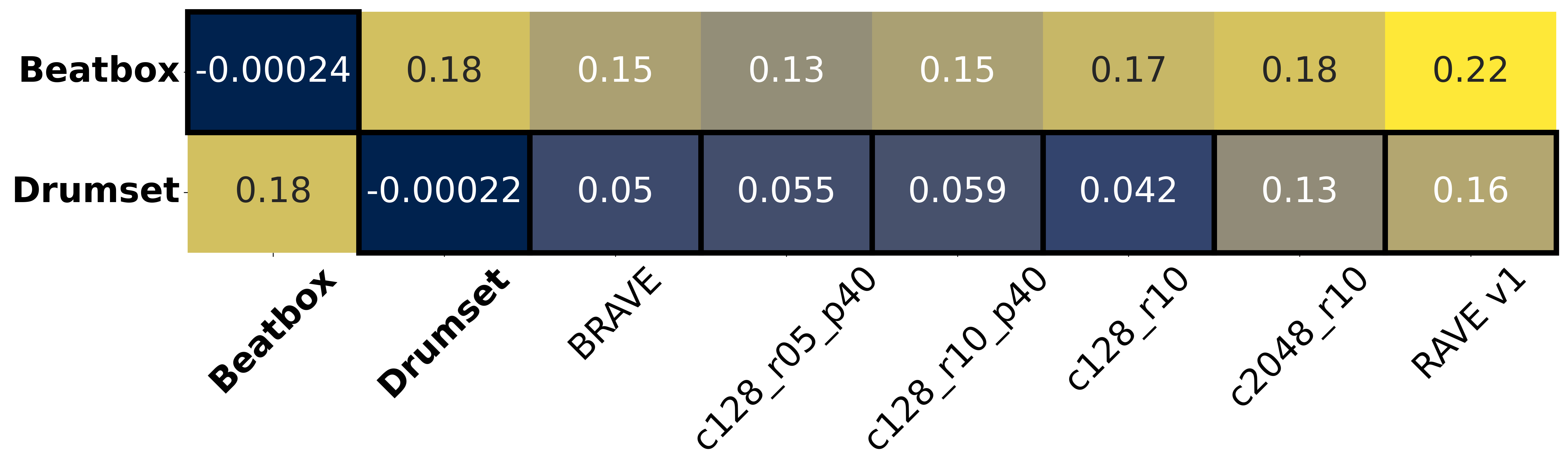}
        \caption*{\textbf{Beatbox} processed by models trained on \textbf{Drumset}}
        \label{fig:sub3}
    \end{subfigure}
    
    \vskip\baselineskip  

    \begin{subfigure}[b]{0.49\textwidth}
        \centering
        \includegraphics[width=\textwidth]{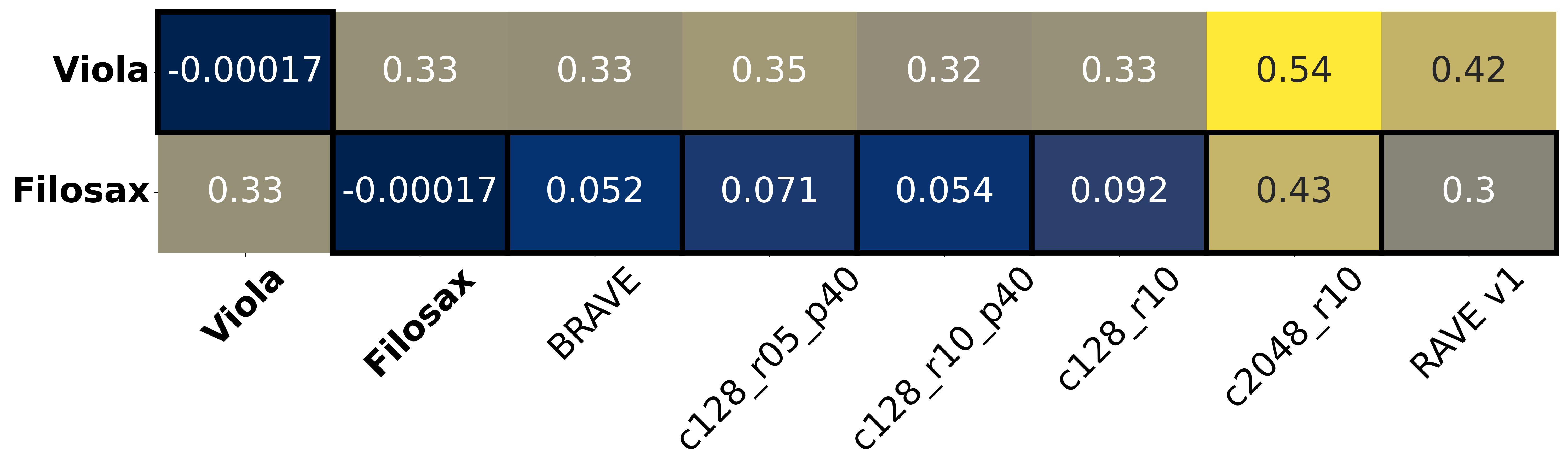}
        \caption*{\textbf{Viola} processed by models trained on \textbf{Filosax}}
        \label{fig:sub2}
    \end{subfigure}
    \hfill
    \begin{subfigure}[b]{0.49\textwidth}
        \centering
        \includegraphics[width=\textwidth]{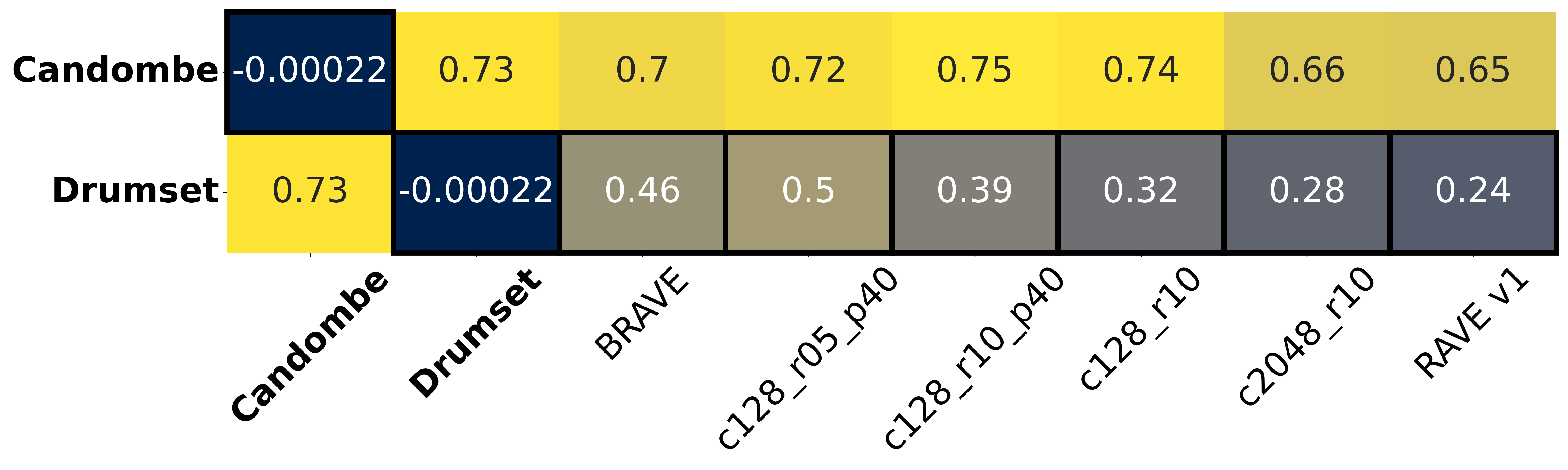}
        \caption*{\textbf{Candombe} processed by models trained on \textbf{Drumset}}
        \label{fig:sub4}
    \end{subfigure}
    
    \caption{We compute the MMD distance between the timbre transfer results of each model, and the input (top-row) and target instrument (bottom-row) datasets. 
    Along the horizontal, entries in bold show dataset cross-similarity and self-similarity. Following this, MMD results for timbre transfer of all model variants. Lower values for the target instrument indicate closer alignment to the target distribution and therefore, a more successful timbre transfer.
    }
    \label{fig:mmd}
\end{figure*}

We perform timbre transfer by running inference over the testing datasets using selected models and storing the results of each model and test set as separate corpora.
Furthermore, we keep the held-out \textbf{Filosax} and \textbf{Drumset} sets as references: we employ them as anchors to evaluate target similarity.
For each timbre transferred corpus, we calculate two MMD distances: one from the test (input instrument) set and the other from the reference set (target instrument). Additionally, we assess the cross-similarity (distance between the test and reference sets) and the self-similarity (distances obtained by sampling from the same test or reference distributions). Successful transfer results should yield MMD scores lower than the cross-similarity, demonstrating that the timbre of the converted audio becomes more aligned with that of the reference dataset.
Results are shown in Figure \ref{fig:mmd}. 

We observe that the MMD scores do not vary significantly across the models. This indicates that the resulting MFCC distributions are not affected by model causality, PQMF attenuation, or compression ratio, suggesting that the timbre modification capabilities of all models are afforded by their receptive field, which even for BRAVE, covers a sizable 0.5s.
The exceptions are the high compression ratio models trained on Filosax, which can be explained due to their inability to render pitched sounds, as discussed in Section \ref{sec:audioquality}.

\subsection{Content Preservation Evaluation}

In this section, we analyze the second requirement for successful timbre transfer: content preservation. We compute a loudness similarity between the test and timbre-transferred corpora, which can account for dynamics, accents, and rhythm information. 
Additionally, we evaluate our models' capability of preserving the fundamental frequency of our melodic test sets account for pitched and melodic content.

Pitch evaluation utilizes pitch tracking computed on input and transferred audio, and evaluates differences using pitch accuracy.
Following \cite{hantrakul_fast_2019}, we use CREPE \cite{kim_crepe_2018} for pitch tracking, which produces pitch estimates at a frame rate of 200Hz.
Results are filtered using a pitch confidence threshold of 0.85 to account for instabilities in pitch estimates.
Accuracy is measured using the overall accuracy metric described by Salamon et al. \cite{salamon2014melody}.
This measures the proportion of frames that are accurately labeled as voiced (i.e., have a pitch confidence above the threshold) and are within 0.5 semitones of the input if the frame is voiced.
Pitch evaluation results are shown in Table \ref{tab:eval_f0}.

Loudness error is the $L_1$ distance between loudness envelopes computed on the input and output audio.
Following Hantrakul et al. \cite{hantrakul_fast_2019}, loudness is calculated by A-Weighting the power spectrum, which models the frequency-dependent hearing sensitivities of the human ear by de-emphasizing low and high frequencies.
A-Weighed power spectrums are then converted to decibel units by log-scaling.
Results are computed using an FFT size of 2048 samples and a hop size of 256 samples.
Loudness $L_1$ distance results are shown in Table \ref{tab:eval_loudness}.

We observe that all content metrics improve with a smaller compression ratio. We attribute this to the higher temporal resolution afforded by all models of $C_r = 128$, which permits them to follow closely fast-varying input signal characteristics related to pitch variations and accents. Interestingly, the BRAVE model outperforms their higher-capacity counterparts on many test sets, with minimal audio quality loss. Being a lower-capacity model, we suspect it learns to rely much more on the encoder to generate plausible output, allowing it to relay more content to the timbre-transferred output. 

\begin{table}[ht]\centering
\scriptsize{
\begin{tabular}{lll}
\hline
 & Svoice & Viola \\
\hline
RAVE v1 & 0.29 & 0.53 \\
\verb|c2048_r10| & 0.28 & 0.53 \\
\verb|c128_r10| & 0.64 & 0.65 \\
\verb|c128_r10_p40| & 0.62 & 0.63 \\
\verb|c128_r05_p40| & 0.63 & 0.67 \\
BRAVE & \textbf{0.68} & \textbf{0.69} \\
\hline
\end{tabular}
\caption{Pitch accuracy evaluated on melodic test sets. Higher values are better.
}\label{tab:eval_f0}
} 
\end{table}

\begin{table}[ht]\centering
\scriptsize{
\begin{tabular}{lllll}
\hline
 & Beatbox & Candombe & Svoice & Viola \\
\hline
RAVE v1 & 14.24 & 13.75 & 33.63 & 24.69 \\
\verb|c2048_r10| & 13.28 & 8.21 & 34.24 & 31.83 \\
\verb|c128_r10| & 9.84 & 4.09 & 12.65 & 9.02 \\
\verb|c128_r10_p40| & 10.97 & 4.08 & 13.68 & 10.02 \\
\verb|c128_r05_p40| & 7.99 & \textbf{4.04} & 12.73 & 7.66 \\
BRAVE & \textbf{7.89} & 4.31 & \textbf{9.94} & \textbf{7.63} \\
\hline
\end{tabular}
\caption{Loudness $L_1$ evaluated on both the percussive and melodic test sets. Lower values are better.}
\label{tab:eval_loudness}
} 
\end{table}

\subsection{A comment on the interaction capabilities of BRAVE}
We deploy several BRAVE models on the demo plugin, which can be downloaded from the accompanying webpage. We play with them using different instruments such as guitar, congas, and voice. The models demonstrate fast response on transients, resulting in very low perceived latency. However, in models like \textbf{Drumset}, the transients appear a bit soft or smeared. Despite this, they perform well in percussive transformations.
For pitched audio, the results are mixed. While the \textbf{Filosax} model also exhibits low latency, it often struggles with correct pitch rendering, which is essential for supporting natural interaction with many instruments. In contrast, other models we test show better performance in this regard; we leave further investigation on pitch performance for future work.

\subsection{Discussion of Results}\label{section:discussion}

\textbf{Data-dependent latency and jitter: }
In NAS systems that leverage convolutional layers, the entire system can be viewed as a complex, non-linear filter where filter coefficients are learned through a data-driven process.
The response of this system, and thereby the overall group delay, is characterized by the underlying convolutional filters.
While the analytic determination of the group delay in such systems is complicated due to non-linearities, we empirically observed that the nature of the data used to train our models had an impact on latency and jitter, consistently measuring lower latency and jitter with the \textbf{Drumset} dataset as compared to the \textbf{Filosax}.
This suggests that signals with strong transients resulted in our models prioritizing temporal accuracy, which contributes heavily to the error in the multi-resolution spectral loss in bands with higher temporal resolution. Some variation of the jitter in the results could be attributed to a different error range of the onset detector between datasets.

Overall, we note that there is a complex relationship between the training data, loss functions, and model latency, which we highlight as an area for future investigation. For example, time-domain audio loss functions may be incorporated to improve temporal precision and phase reconstruction of the model \cite{webber_autovocoder_2023}; however, this may be incompatible with the variational autoencoder objective.

\textbf{Timbre transfer affordances:}
Our timbre transfer test results indicate that timbre modification can be afforded by a large receptive field, evidenced by the similar MMD score between BRAVE and RAVE. Furthermore, content rendering improves when working at smaller compression ratios. This is in line with the typical timbre transfer design assumptions that assume that musical content varies over time while timbre is a global attribute.
Finally, it is surprising to see similar performance between our lower-capacity BRAVE model and its low-compression ratio counterparts. This suggests that (1) further performance gains can be obtained by a careful revision of BRAVE's structure towards a smaller model, and (2) there may be potential for increasing audio fidelity in bigger models.

\textbf{Design challenges for low latency NAS:}
Designing NAS algorithms for low latency requires putting this issue at the center of the design problem, and requires a careful inspection of the audio representations and the architecture to ensure these do not increase response time. We find that a representation learned with a VAE, rendered as a \textit{time-varying trajectory} progressively constructed over short audio windows in a temporally causal manner by a compressing encoder and aggregated by a decoder with a considerable receptive field, works well for our timbre transfer task.

We believe this approach can work for other low-latency interactive NAS tasks, provided all modules within the network are designed to support such latent trajectories. For instance, a low-latency re-design of RAVE's noise generator for low compression ratio operation could improve BRAVE's audio quality. One possibility would be the addition of low-latency, differentiable Infinite Impulse Response filters \cite{yu_differentiable_2024} controlled by an additional decoder submodule with an appropriate receptive field.
We expect our design experience can guide other researchers in the design of complex low-latency NAS algorithms, while leveraging recent inference tools that can simplify implementation \cite{ackva_anira_2024,chowdhury_rtneural_2021}.

\section{CONCLUSION}

In this paper, we advocated for designing interactive NAS models from the ground up, considering latency at each step.
To support this design process we presented an overview of common sources of latency and jitter, along with a method for empirically measuring it.
We find that RAVE, a model with extensive interactive applications, incurs a level of latency that makes it unsuitable for audio-driven interaction with traditional musical instruments.

Our redesign process puts latency at the center of the problem and allows us to design architectures that are suitable for real-time interaction with musical instruments. This can enable timbre transfer in use cases that as of today have been relegated to production environments or specially designed musical interfaces.

Our low-latency timbre transfer system holds the potential to foster intimate control of instruments and interfaces with an extended timbral palette. However, its data-dependent performance may limit interaction possibilities, and further work should look into how its affordances are affected by data and its corresponding latent representations. 


\section{ACKNOWLEDGMENT}
This work is supported by the EPSRC UKRI Centre for Doctoral Training in Artificial Intelligence and Music (EP/S022694/1). This research utilized Queen Mary’s Apocrita HPC facility, supported by QMUL Research-IT. http://doi.org/10.5281/zenodo.438045.

\bibliography{main}
\bibliographystyle{IEEEtran}

\end{document}